\documentclass[aps,showpacs,preprintnumbers,amsmath,amssymb]{revtex4}

\usepackage{float}
\usepackage{graphics,epsfig}
\usepackage{graphicx}
\usepackage{dcolumn}
\usepackage{bm}
\newcommand{\no}{\nonumber}

\begin{document}
\baselineskip=0.8 cm
\title{{\bf Hawking radiation in a Rotating Kaluza-Klein Black Hole
with Squashed Horizons}}
\author{Songbai Chen}
\email{csb3752@163.com} \affiliation{Department of Physics, Fudan
University, Shanghai 200433, P. R. China
 \\ Institute of Physics and  Department of Physics,
Hunan Normal University,  Changsha, Hunan 410081, P. R. China }

\author{Bin Wang}
\email{wangb@fudan.edu.cn} \affiliation{Department of Physics,
Fudan University, Shanghai 200433, P. R. China}

\author{ Rukeng Su}
\email{rksu@fudan.ac.cn}
 \affiliation{China Center of Advanced Science and Technology (World Laboratory),
P.O.Box 8730, Beijing 100080, People¡¯s Republic of China
\\ Department of Physics, Fudan University, Shanghai 200433, P. R. China}

\vspace*{0.2cm}
\begin{abstract}
\baselineskip=0.6 cm
\begin{center}
{\bf Abstract}
\end{center}

We explore the signature of the extra dimension in the Hawking
radiation in a rotating Kaluza-Klein black hole with squashed
horizons. Comparing with the spherical case, we find that the
rotating parameter brings richer physics. We obtain the
appropriate size of the extra dimension which can enhance the
Hawking radiation and may open a window to detect the extra
dimensions.

\end{abstract}

 \maketitle
\newpage
\section{Introduction}

String theory is believed as the promising candidate for the
unified theory of everything, which predicts the existence of the
extra dimension. It is of great interest to study whether the
extra dimension can be observed. This can present the signature of
string and the correctness of string theory.

A great deal of effort has been expanded for the detection of the
extra dimension. One among them is the study of the perturbation
around braneworld black holes. It has been argued that the extra
dimension can imprint in the wave dynamics in the branworld black
hole background \cite{Shen, Abdalla,Chen,Kanti}. Another chief
possibility of observing the extra dimension is the spectrum of
Hawking radiation which is expected to be detected in particle
accelerator experiments \cite{5, 6, 7, 8, 9, 10, 11, 12, 13}.
Recently through the study of Hawking radiation from squashed
Kaluza-Klein (KK) black holes \cite{IM}, it was argued that the
luminosity of Hawking radiation can tell us the size of the extra
dimension which opens a window to observe extra dimensions
\cite{sq1hw}.

Besides the charged static KK black hole with squashed horizon found
in five-dimensional Einstein-Maxwell theory \cite{IM}, other
solutions of squashed KK black hole \cite{sq2,sq3,sq4} have also
been obtained subsequently. A nice review on the KK black hole
including its stability, phase diagram and thermodynamics can be
found in \cite{TVN} and references therein. Recently, an extension
has been made to the rotating case with two equal angular momenta in
Einstein theory with zero cosmological constant\cite{sq5}. The
spacetime of the rotating KK black hole solution with squashed
horizons is geodesic complete and free of naked singularities. It
has the similar topology and asymptotical structure to that of the
static squashed KK black hole, but with richer physical properties.
In this paper we are going to study the Hawking radiation in the
rotating squashed KK black hole background. We will calculate the
greybody factor of scalar particles propagating in this rotating
squashed KK black hole and show the rich physics brought by the
rotating parameter $a$.

\section{Master equation in the squashed Kerr black
hole}

The five-dimensional rotating squashed KK black hole with two
equal angular momenta is described by\cite{sq5}
\begin{eqnarray}
ds^2=-dt^2+\frac{\Sigma_0}{\Delta_0}k(r)^2dr^2+\frac{r^2+a^2}{4}
[k(r)(\sigma^2_1+\sigma^2_2)+\sigma^2_3]+\frac{M}{r^2+a^2}(dt-\frac{a}{2}\sigma_3)^2,
\label{metric0}
\end{eqnarray}
with
\begin{eqnarray}
\sigma_1&=&-\sin{\psi} d\theta+\cos{\psi} \sin{\theta}d\phi,\nonumber\\
\sigma_2&=&\cos{\psi} d\theta+\sin{\psi} \sin{\theta}d\phi,\nonumber\\
\sigma_3&=&d\psi+\cos{\theta}d\phi,
\end{eqnarray}
where $0<\theta<\pi$, $0<\phi<2\pi$ and $0<\psi<4\pi$. The
parameters are given by
\begin{eqnarray}
\Sigma_0&=&r^2(r^2+a^2),\nonumber\\
\Delta_0&=&(r^2+a^2)^2-Mr^2,\nonumber\\
k(r)&=&\frac{(r^2_{\infty}-r^2_+)(r^2_{\infty}-r^2_-)}{(r^2_{\infty}-r^2)^2}.
\end{eqnarray}
Here $M$ and $a$ correspond to mass and angular momenta,
respectively. $r=r_+$ and $r=r_-$ are outer and inner horizons of
the black hole and they relate to $M,a$ by $a^4=(r_+r_-)^2,
M-2a^2=r^2_++r^2_-$. $r_{\infty}$ corresponds to the spatial
infinity. In the parameter space $0 < r_-\leq r_+ < r_{\infty}$,
$r$ is restricted within the range $0<r<r_{\infty}$. The shape of
black hole horizon is deformed by the parameter $k(r_+)$.

In the metric (1), the intrinsic singularity is just the one at
$r=0$, while $r_{\pm}$ and $r_{\infty}$ are coordinate
singularities. This can be seen by introducing a new radial
coordinate $\rho$ as \cite{sq5}
\begin{eqnarray}
\rho=\rho_0\frac{r^2}{r^2_{\infty}-r^2},\label{p}
\end{eqnarray}
with
\begin{eqnarray}
\rho^2_0&=&\frac{k_0}{4}(r^2_{\infty}+a^2),\nonumber \\
k_0&=&k(r=0)=\frac{(r^2_{\infty}+a^2)^2-Mr^2_{\infty}}{r^4_{\infty}}.
\end{eqnarray}
At the spatial infinity $r\rightarrow r_{\infty},
\rho\rightarrow\infty$. Thus in the new coordinate, $\rho$ varies
from $0$ to $\infty$ when $r$ changes from $0$ to $r_{\infty}$.
The metric (\ref{metric0}) can be rewritten as
\begin{eqnarray}
ds^2=-\frac{(r^2_{\infty}+a^2)^4}{4\rho^2_0r^6_{\infty}}d\tau^2+Ud\rho^2+R^2(\sigma^2_1+\sigma^2_2)
+W^2\sigma^2_3+V\bigg[\frac{(r^2_{\infty}+a^2)^2}{2\rho_0r^3_{\infty}}d\tau-\frac{a}{2}\sigma_3\bigg]^2,
\label{metric}
\end{eqnarray}
where we have defined the proper time
$\tau=2\rho_0r^3_{\infty}/(r^2_{\infty}+a^2)^2t$ for the observer
at infinity. The quantities $K$, $V$, $W$, $R$ and $U$ are
functions of $\rho$
\begin{eqnarray}
K^2&=&\frac{\rho+\rho_0}{\rho+\frac{a^2}{r^2_{\infty}+a^2}\rho_0},\nonumber\\
V&=&\frac{M}{r^2_{\infty}+a^2}K^2,\nonumber\\
W^2&=&\frac{r^2_{\infty}+a^2}{4K^2}=\frac{M}{4V},\nonumber\\
R^2&=&\frac{(\rho+\rho_0)^2}{K^2},\nonumber\\
U&=&(\frac{r^2_{\infty}}{r^2_{\infty}+a^2})^2\times\frac{\rho^2_0}{W^2-\frac{r^2_{\infty}}{4}
\frac{\rho}{\rho+\rho_0}V}.
\end{eqnarray}
The Hawking temperature for this rotating squashed KK black hole
is given by
\begin{eqnarray}
T_H=\frac{(r^2_+-r^2_-)r_+}
{2\pi(r^2_++a^2)^2}\sqrt{1-\frac{r^2_+}{r^2_{\infty}}}
\bigg[\frac{r^2_{\infty}+a^2}{r^2_{\infty}-r^2_-}\bigg]^{3/2}
=\frac{\rho_+-\rho_-}
{4\pi(\rho_++\frac{a^2}{r^2_{\infty}+a^2}\rho_0)^2}\sqrt{\frac{\rho_+}{\rho_++\rho_0}},\label{TH}
\end{eqnarray}
where $\rho_{\pm}=\rho_0\frac{r^2_{\pm}}{r^2_{\infty}-r^2_{\pm}}$
are the outer and inner horizons of the black hole in the new
coordinate. The angular velocity at the event horizon is
\begin{eqnarray}
\Omega_H=\frac{a
K(\rho_+)^2(r^2_{\infty}+a^2)}{\rho_0r^3_{\infty}},
\end{eqnarray}

The determinant of the metric is
\begin{eqnarray}
g=-\frac{(r^2_{\infty}+a^2)^2(\rho+\rho_0)^4} {4
r^2_{\infty}K^4}\sin^2{\theta},
\end{eqnarray}
and non-zero metric coefficients read
\begin{eqnarray}
&&g^{00}=-\frac{4\rho^2_0r^6_{\infty}}{(r^2_{\infty}+a^2)^4}\bigg[\frac{\Delta
+M(r^2_{\infty}+a^2)^2(\rho+\rho_0)^2/(4\rho^2_0r^4_{\infty}K^2)}{\Delta
}\bigg],\nonumber\\
&&g^{11}=\frac{K^2\Delta
}{(\rho+\rho_0)^2},\;\;\;\;\;\;\;\;\;\;\;\;\;\;\;\;\;\;\;\;\;\;\;\;\;\;\;\;
g^{22}=\frac{K^2}{(\rho+\rho_0)^2},\nonumber\\
&&g^{33}=\frac{K^2}{(\rho+\rho_0)^2\sin^2{\theta}},\;\;\;\;\;\;\;\;\;\;\;\;\;\;\;\;\;\;\;
g^{34}=-\frac{K^2\cos{\theta}}{(\rho+\rho_0)^2\sin^2{\theta}},
\nonumber\\
&&g^{04}=-\frac{Ma(\rho+\rho_0)^2}{\rho_0r_{\infty}(r^2_{\infty}+a^2)\Delta},\nonumber\\
&&g^{44}=\frac{K^2\cos^2{\theta}}{(\rho+\rho_0)^2\sin^2{\theta}}+
\frac{(r^2_{\infty}+a^2)(\rho+\rho_0)[(r^2_{\infty}+a^2)\rho+a^2\rho_0-M(\rho+\rho_0)]}{\rho^2_0r^4_{\infty}\Delta},
\end{eqnarray}
where
\begin{eqnarray}
\Delta=\frac{[(r^2_{\infty}+a^2)\rho+a^2\rho_0]^2-Mr^2_{\infty}\rho(\rho+\rho_0)}
{(r^2_{\infty}+a^2)^2-Mr^2_{\infty}}.
\end{eqnarray}
We note that $g^{03}$ vanishes, which is different from that in
the usual five dimensional Kerr black hole with different angular
momenta. This implies that there exists some special properties in
such spacetime.

The wave equation for the massless scalar field
$\Phi(t,r,\theta,\phi,\psi)$ in the background (\ref{metric})
obeys
\begin{eqnarray}
\frac{1}{\sqrt{-g}}\partial_{\mu}(\sqrt{-g}g^{\mu\nu}\partial_{\nu})
\Phi(t,r,\theta,\phi,\psi)=0.\label{WE}
\end{eqnarray}
Taking the ansatz $\Phi(t,r,\theta,\phi,\psi)=e^{-i\omega
t}R(\rho)e^{i m\phi+i\lambda \psi}S(\theta)$, where $S(\theta)$ is
the so-called spheroidal harmonics, we can obtain the equation
\begin{eqnarray}
\frac{1}{\sin{\theta}}\frac{d}{d\theta}\bigg[\sin{\theta}
\frac{d}{d\theta}\bigg]S(\theta)
-\bigg[\frac{(m-\lambda\cos{\theta})^2}{\sin^2{\theta}}-E_{lm\lambda}\bigg]S(\theta)=0,\label{angd}
\end{eqnarray}
for angular part. Obviously, this angular equation is independent of
the rotating parameter $a$, and is exactly identical to that in the
static squashed KK black hole spacetime \cite{sq1hw}. This is not
surprising and in \cite{sq6} the same angular equation as that in
the Schwarzschild spacetime was also obtained in the
five-dimensional Kerr black hole with two equal rotational
parameters. The eigenvalue of the angular equation (\ref{angd}) is
$E_{lm\lambda}=l(l+1)-\lambda^2$. The radial part reads
\begin{eqnarray}
\frac{d}{d\rho}\bigg[\Delta\frac{d R(\rho)}{d\rho}\bigg]
+\bigg[\frac{\tilde{K}^2}{\Delta}+\Lambda-E_{lm\lambda}\bigg]R(\rho)=0,\label{radial}
\end{eqnarray}
with
\begin{eqnarray}
&&\tilde{K}^2=\frac{Mr^2_{\infty}(\rho+\rho_0)^4}{K^4(r^2_{\infty}+a^2)^2}
\bigg[\omega-\frac{\lambda a
K^2(r^2_{\infty}+a^2)}{\rho_0r^3_{\infty}}\bigg]^2,
\\
&&\Lambda=\frac{4\rho^2_0r^6_{\infty}(\rho+\rho_0)^2}{K^2(r^2_{\infty}+a^2)^4}\omega^2-
\frac{4\lambda^2(\rho+\rho_0)^2}{r^2_{\infty}+a^2}.
\end{eqnarray}
By solving equation (\ref{radial}), one can obtain the radial part
of the wave-function $R(\rho)$ , and then compute the absorption
probability $|\mathcal{A}_{lm\lambda}|^2$ and study the Hawking
radiation of a scalar particle propagating in the black hole
spacetime.

\section{Greybody Factor in the Low-Energy Regime}

In this section, we will obtain the solution to the field equation
by employing matching techniques on expressions valid in the near
horizon $(\rho\sim \rho_+)$ and far field $(\rho\gg\rho_+)$
regimes in the low energy and low angular momentum limit.

Let us first focus on the near-horizon regime. In order to express
equation (\ref{radial}) into the form of a known differential
equation,  we make the following change of variable
\begin{eqnarray}
\rho\rightarrow f=\frac{\Delta K^4}{(\rho+\rho_0)^2}\Rightarrow
\frac{d f}{d\rho}=(1-f)\frac{A\;K^2}{(\rho+\rho_0)},
\end{eqnarray}
where
\begin{eqnarray}
A=1-\frac{\rho_0 r^2_{\infty}a^2}{\rho
(r^4_{\infty}-a^4)-a^4\rho_0}.
\end{eqnarray}
The equation (\ref{radial}) near the horizon $(\rho\sim \rho_+)$
can be expressed as
\begin{eqnarray}
f(1-f)\frac{d^2R(f)}{d f^2}+(1-D_*f)\frac{d R(f)}{d f}
+\bigg\{\frac{K^2_*}{A(\rho_+)^2(1-f)f}
-\frac{E_{lm\lambda}-\Lambda(\rho_+)}{A(\rho_+)^2(1-f)}\bigg\}R(f)=0,\label{r1}
\end{eqnarray}
where
\begin{eqnarray}
K_*&=&\sqrt{1+\frac{\rho_0}{\rho_+}}\bigg[\rho_++\frac{a^2}{r^2_{\infty}+a^2}\rho_0\bigg]\bigg[\omega-\frac{a
\lambda
K(\rho_+)^2(r^2_{\infty}+a^2)}{\rho_0r^3_{\infty}}\bigg],\nonumber\\
D_*&=&2-\frac{1}{A(\rho_+)}-\frac{(\rho_++\rho_0)A'(\rho_+)}{K(\rho_+)^2A(\rho_+)^2}.
\end{eqnarray}
Here $A'(\rho_+)$ denotes the derivative of $A$ with respect to
$\rho$ at the outer horizon $\rho=\rho_+$. Redefining the field
$R(f)=f^{\alpha}(1-f)^{\beta}F(f)$, we can write equation
(\ref{r1}) into the form of hypergeometric equation
\begin{eqnarray}
f(1-f)\frac{d^2F(f)}{d f^2}+[c-(1+a_1+b)f]\frac{d F(f)}{d f}-a_1b
F(f)=0,\label{r2}
\end{eqnarray}
with
\begin{eqnarray}
a_1=\alpha+\beta+D_*-1,\;\;\;\;\;\;\;\;\;\;
b=\alpha+\beta,\;\;\;\;\;\;\;\;\;\;\;\;\; c=1+2\alpha.
\end{eqnarray}
Due to the constraint from coefficient of $F(f)$, the power
coefficients $\alpha$ and $\beta$ must satisfy the second-order
algebraic equations
\begin{eqnarray}
\alpha^2+\frac{K^2_*}{A(\rho_+)^2}=0,
\end{eqnarray}
and
\begin{eqnarray}
\beta^2+\beta(D_*-2)+\frac{K^2_*}{A(\rho_+)^2}-
\frac{E_{lm\lambda}-\Lambda(\rho_+)}{A(\rho_+)^2}=0,
\end{eqnarray}
respectively. Solving these two equations, we obtain the solutions
for the parameter $\alpha$ and $\beta$
\begin{eqnarray}
&&\alpha_{\pm}=\pm \frac{iK_*}{A(\rho_+)},\\
&&\beta_{\pm}=\frac{1}{2}\bigg[(2-D_*)\pm\sqrt{(D_*-2)^2-\frac{4K^2_*}{A(\rho_+)^2}
+\frac{4(E_{lm\lambda}-\Lambda(\rho_+))}{A(\rho_+)^2}} \;\bigg].
\end{eqnarray}
The general solution of the master equation (\ref{radial}) near
the horizon can be expressed as
\begin{eqnarray}
R_{NH}(f)=A_-f^{\alpha}(1-f)^{\beta}F(a_1,b,c;
f)+A_+f^{-\alpha}(1-f)^{\beta}F(a_1-c+1,b-c+1,2-c; f),\label{s0}
\end{eqnarray}
where $A_{\pm}$ are arbitrary constants. Near the horizon,
$\rho\rightarrow\rho_+$, and $f\rightarrow 0$, the solution
(\ref{s0}) can be reduced to
\begin{eqnarray}
R_{NH}(f)=A_-f^{\pm iK_*/A(\rho+)}+A_+f^{\mp
iK_*/A(\rho+)}=A_-e^{\pm i\mathcal{K}y}+A_+e^{\mp
i\mathcal{K}y},\label{s1}
\end{eqnarray}
with
\begin{eqnarray}
\mathcal{K}=\sqrt{1+\frac{\rho_0}{\rho_+}}\bigg[\rho_++\frac{a^2}{r^2_{\infty}+a^2}\rho_0\bigg]^2\bigg[\omega-\frac{a
\lambda K(\rho_+)^2(r^2_{\infty}+a^2)}{\rho_0r^3_{\infty}}\bigg]
\end{eqnarray}
where $y$ is the tortoise-like coordinate and can be expressed as
\begin{eqnarray}
y=\frac{K(\rho_+)^2\ln{f}}{A(\rho_+)(\rho_++\rho_0)}.
\end{eqnarray}
In the limit $\rho\rightarrow \rho_+$, it becomes identical to the
tortoise coordinate $\rho_*$ defined by
$d\rho_*/d\rho=K^4/f(\rho+\rho_0)^2$ as in \cite{sq1hw}. Thus, the
factors $f^{\pm iK_*/A(\rho+)}$ in the near-horizon asymptotic
solution can be reduced to $e^{\pm i\mathcal{K}y}$, which
describes an outgoing and incoming free waves, respectively. In
terms of the boundary condition that no outgoing mode exists near
the horizon, we choose $\alpha=\alpha_-$ and $A_+=0$, which
results in the asymptotic solution near horizon
\begin{eqnarray}
R_{NH}(f)=A_-f^{\alpha}(1-f)^{\beta}F(a_1, b, c; f).
\end{eqnarray}
Moreover, the above boundary condition also demands that near the
horizon the hypergeometric function $F(a_1, b, c; f)$ must be
convergent, i.e. $Re(c-a_1- b)> 0$, which implies that we must
choose $\beta=\beta_-$.

In order to match the near horizon and far field solutions in the
intermediate zone, we must stretch the near horizon solution to
the large value of the radial coordinate. As done in
Ref.\cite{Haw3}, at first we change the argument of the
hypergeometric function of the near-horizon solution from $f$ to
$1-f$ by using the relation
\begin{eqnarray}
R_{NH}(f)&=&A_-f^{\alpha}(1-f)^{\beta}\bigg[\frac{\Gamma(c)\Gamma(c-a_1-b)}{\Gamma(c-a_1)\Gamma(c-b)}
F(a_1, b, a_1+b-c+1; 1-f)\nonumber\\
&+&(1-f)^{c-a_1-b}\frac{\Gamma(c)\Gamma(a_1+b-c)}{\Gamma(a_1)\Gamma(b)}
F(c-a_1, c-b, c-a_1-b+1; 1-f)\bigg].\label{r2}
\end{eqnarray}
In the limit $\rho\gg\rho_+$, the function $(1-f)$ can be
approximated by
\begin{eqnarray}
1-f\simeq \frac{M(r^2_{\infty}-a^2)}{4\rho_0 r^2_{\infty}\rho},
\end{eqnarray}
and the near-horizon solution (\ref{r2}) can be simplified further
to
\begin{eqnarray}
R_{NH}(\rho)\simeq
A_1\rho^{-\beta}+A_2\rho^{\;\beta+D_*-2}\label{rn2},
\end{eqnarray}
with
\begin{eqnarray}
A_1=A_-\bigg[\frac{M(r^2_{\infty}-a^2)}{4\rho_0
r^2_{\infty}}\bigg]^{\beta}
\frac{\Gamma(c)\Gamma(c-a_1-b)}{\Gamma(c-a_1)\Gamma(c-b)},\label{rn3}
\end{eqnarray}
\begin{eqnarray}
A_2=A_-\bigg[\frac{M(r^2_{\infty}-a^2)}{4\rho_0
r^2_{\infty}}\bigg]^{-(\beta+D_*-2)}\frac{\Gamma(c)\Gamma(a_1+b-c)}{\Gamma(a_1)\Gamma(b)}.\label{rn4}
\end{eqnarray}

Now, let us turn to the far field region $(\rho\rightarrow
\infty)$, where the equation (\ref{radial}) reduces to
\begin{eqnarray}
\frac{d^2R_{FF}(\rho)}{d
\rho^2}+\frac{2}{\rho}\frac{dR_{FF}(\rho)}{d
\rho}+\bigg[\tilde{\omega}^2-\frac{E_{lm\lambda}}{\rho^2}\bigg]R_{FF}(\rho)=0,
\end{eqnarray}
with
\begin{eqnarray}
\tilde{\omega}^2=\frac{4\rho^2_0r^6_{\infty}\omega^2}{(r^2_{\infty}+a^2)^4}-\frac{4\lambda^2}{r^2_{\infty}+a^2}
+\frac{Mr^2_{\infty}}{(r^2_{\infty}+a^2)^2}\bigg[\omega-\frac{a\lambda(r^2_{\infty}+a^2)}{\rho_0
r^3_{\infty}}\bigg]^2.\label{ws0}
\end{eqnarray}
Obviously, it is a Bessel equation. Thus, the general solution of
radial master equation (\ref{radial}) in the far field region can
be expressed as
\begin{eqnarray}
R_{FF}(\rho)=\frac{1}{\sqrt{\rho}}\bigg[B_1J_{\nu}(\tilde{\omega}\rho)+B_2Y_{\nu}
(\tilde{\omega}\rho)\bigg],\label{rf}
\end{eqnarray}
where $J_{\nu}(\tilde{\omega}\rho)$ and
$Y_{\nu}(\tilde{\omega}\rho)$ are the first and second kind Bessel
functions, $\nu=\sqrt{E_{lm\lambda}+1/4}$. $B_1$ and $B_2$ are
integration constants. In the limit $\rho\rightarrow 0$,
$R_{FF}(\rho)$ in equation (\ref{rf}) becomes
\begin{eqnarray}
R_{FF}(\rho)\simeq\frac{B_1(\frac{\tilde{\omega}\rho}{2})^{\nu}}{\sqrt{\rho}\;\Gamma(\nu+1)}
-\frac{B_2\Gamma(\nu)}{\pi
\sqrt{\rho}\;(\frac{\tilde{\omega}\rho}{2})^{\nu}}.
\end{eqnarray}
Comparing it with equation (\ref{rn2}), we can obtain two
relations between $A_1$ and $B_1,\;B_2$ in the limit $\omega
\rho_+\ll1$. Employing equations (\ref{rn3}) and (\ref{rn4}) and
removing $A_-$, we find the constraint for $B_1,\; B_2$
\begin{eqnarray}
B\equiv\frac{B_1}{B_2}&=&-\frac{1}{\pi}\bigg[\frac{8\rho_0
r^2_{\infty}}{M\tilde{\omega}
(r^2_{\infty}-a^2)}\bigg]^{\sqrt{4E_{lm\lambda}+1}}\sqrt{E_{lm\lambda}+1/4}\nonumber\\
&\times& \frac{\Gamma^2(\sqrt{E_{lm\lambda}+1/4})
\Gamma(c-a_1-b)\Gamma(a_1)\Gamma(b)}{\Gamma(a_1+b-c)\Gamma(c-a_1)\Gamma(c-b)}.
\label{BB}
\end{eqnarray}
In the asymptotic region $\rho\rightarrow \infty$, the solution in
the far field can be expressed as
\begin{eqnarray}
R_{FF}(\rho)\simeq
\frac{B_1+iB_2}{2\sqrt{2\pi\tilde{\omega}}\rho}e^{-i\tilde{\omega}\rho}+
\frac{B_1-iB_2}{2\sqrt{2\pi\tilde{\omega}}\rho}e^{i\tilde{\omega}\rho}=
A^{(\infty)}_{in}\frac{e^{-i\tilde{\omega}\rho}}{\rho}+A^{(\infty)}_{out}\frac{e^{i\tilde{\omega}\rho}}{\rho}.\label{rf6}
\end{eqnarray}
We need to point out that only when the condition
$\tilde{\omega}>0$ is satisfied, the solution (\ref{rf6}) denotes
an incoming and an outgoing spherical waves for large distances
from the black hole.

The absorption probability can be calculated by
\begin{eqnarray}
|\mathcal{A}_{lm\lambda}|^2=1-\bigg|\frac{A^{(\infty)}_{out}}{A^{(\infty)}_{in}}\bigg|^2
=1-\bigg|\frac{B-i}{B+i}\bigg|^2=\frac{2i(B^*-B)}{BB^*+i
(B^*-B)+1}.\label{GFA}
\end{eqnarray}
Combining the above result and the expression $B$ given in
equation (\ref{BB}), we can analyse the properties of absorption
probability for the massless scalar field in a rotating squashed
KK black hole background in the low-energy and low-angular
momentum limits.

\begin{figure}[ht]
\begin{center}
\includegraphics[width=8.5cm]{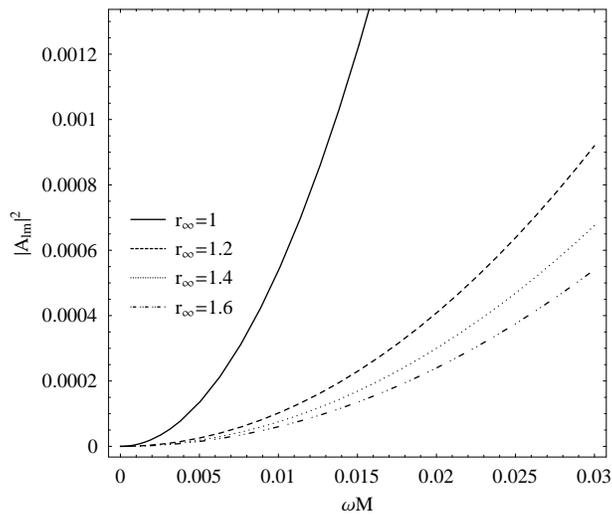}
\caption{Absorption probability $|\mathcal{A}_{lm\lambda}|^2$ of
 scalar particles propagating in the rotating squashed Kaluza-Klein black
hole spacetime, for fixed $l=0$, $\lambda=0$, $a=0.2$ and
different $r_{\infty}$.}
\end{center}
\label{fig1}
\end{figure}

In Fig.(1), we plot the absorption probability for the first
partial waves ($l=0, \lambda=0$) by fixing $a=0.2$. One can easily
see that the absorption probability decreases with the increase of
the parameter $r_{\infty}$, which is similar to that in the
nonrotating case shown in \cite{sq1hw} where Eq.(31) can be
written as $|\mathcal{A}|^2=\frac{\omega^2
r_{\infty}r^3_+}{r^2_{\infty}-r^2_+}$ decreasing with the increase
of $r_{\infty}$. The main reason behind this phenomenon is that
the larger $r_{\infty}$ yields the lower peak of the effective
potential so that more radiation can be transmitted to the
infinity.

\begin{figure}[ht]
\begin{center}
\includegraphics[width=8.2cm]{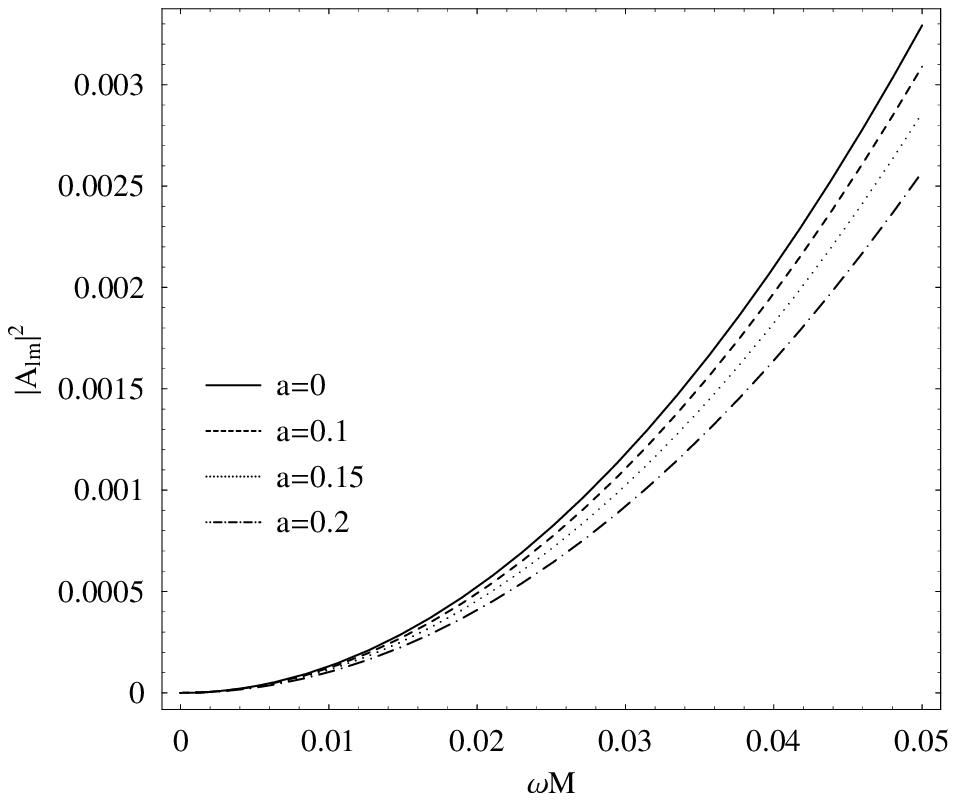}\;\;\;\;\includegraphics[width=8.2cm]{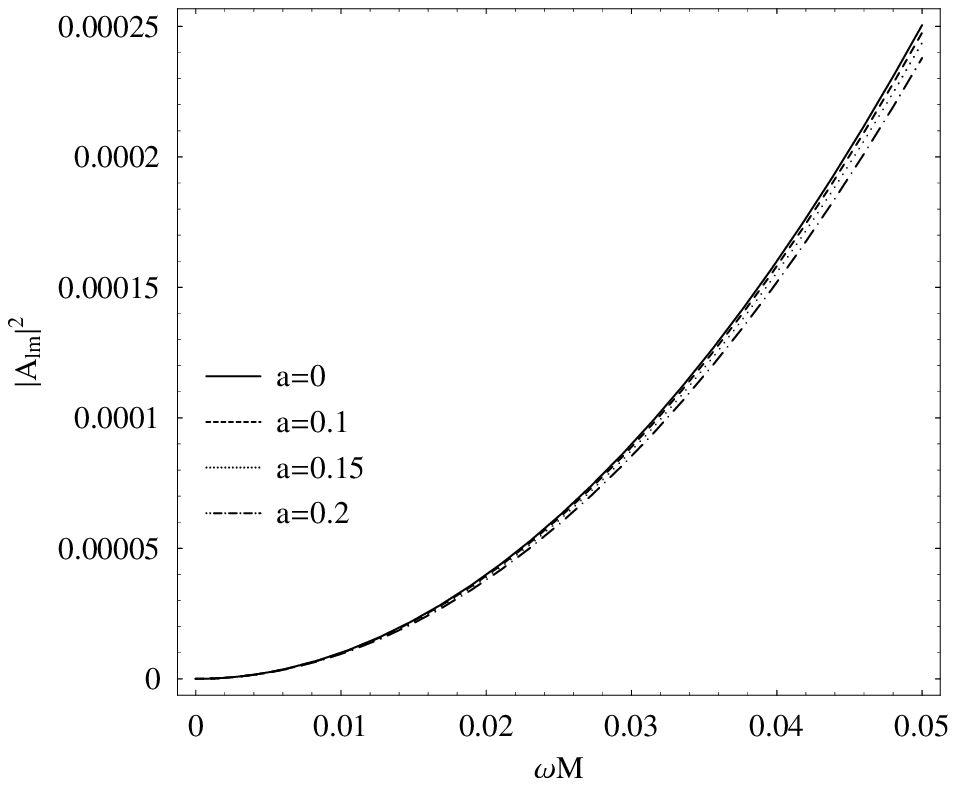}\;\;\;\;\;
\caption{Absorption probability $|\mathcal{A}_{lm\lambda}|^2$ (the
left $r_{\infty}=1.2$ and the right $r_{\infty}=5$) of
 scalar particles propagating in the rotating squashed Kaluza-Klein black
hole spacetime, for fixed $l=0$, $\lambda=0$ and different $a$.  }
 \end{center}
 \label{fig2}
\end{figure}

In figure (2) we show the dependence of the absorption probability
on the angular momentum parameter. We see that for fixed
$r_{\infty}$, the absorption probability decreases with the
increase of the angular moment parameter $a$. The suppression of
the absorption due to the increase of $a$ is very similar to that
of the Reissner-Nordstrom black hole when the value of $q$
increases \cite{RN}. For the fixed $r_{\infty}$, we observed that
for larger $a$, the potential peak becomes lower, which allows
more radiation to leak to the infinity. The difference of the
absorption probability due to different $a$ becomes smaller when
the value of $r_{\infty}$ increases. This is because that big
$r_{\infty}$ causes the spectrum to be suppressed as observed in
Fig.1.

\begin{figure}[ht]
\begin{center}
\includegraphics[width=8cm]{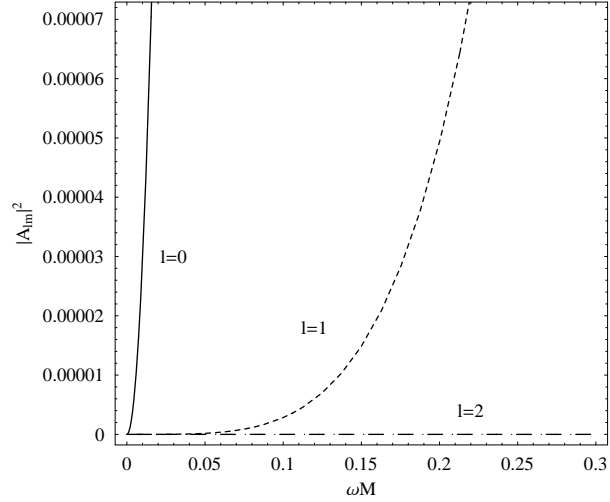}
\caption{Absorption probability $|\mathcal{A}_{lm\lambda}|^2$  of
scalar particles propagating in the rotating squashed Kaluza-Klein
black hole spacetime, for fixed $r_{\infty}=2$, $\lambda=0$,
$a=0.2$ and different $l$.}
\end{center}
\label{fig3}
\end{figure}
\begin{figure}[ht]
\begin{center}
\includegraphics[width=8.cm]{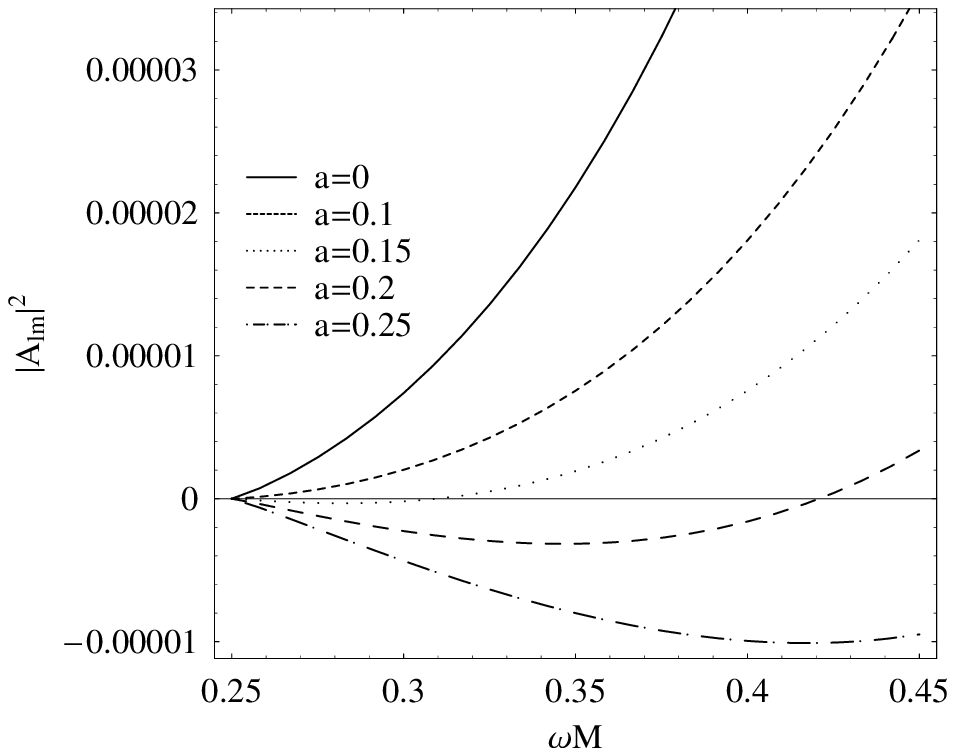}\;\;\;\;\;\includegraphics[width=8.2cm]{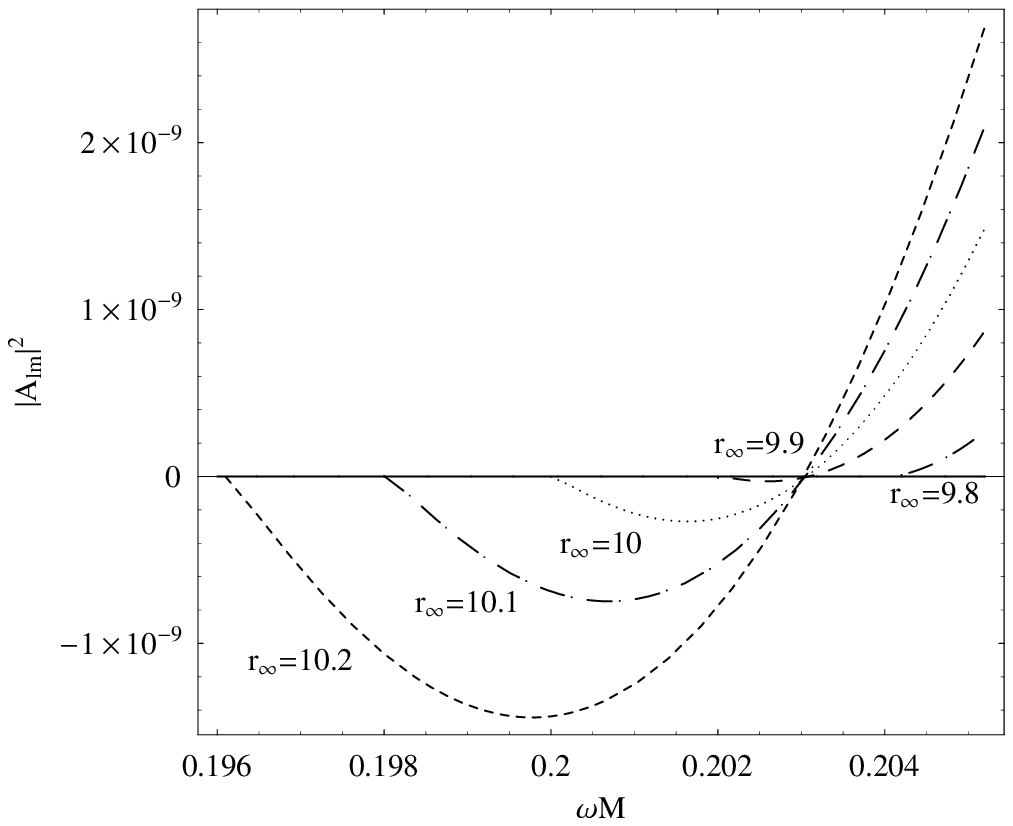}
\caption{Absorption probability $|\mathcal{A}_{lm\lambda}|^2$ of
scalar particles propagating in the rotating squashed Kaluza-Klein
black hole spacetime, for $l=1$, $\lambda=1$. The left is for
fixed $r_{\infty}=5$ and different $a$,  and the right is for
fixed $a=0.1$ and different $r_{\infty}$.}
\end{center}
\label{fig4}
\end{figure}

Fig.3 shows the dependence of the absorption probability on the
angular index $l$. We see the dominance of the low angular index
spectrum over others. This property has also been observed in
\cite{Haw3}. In the low energy approximation, (47) in \cite{Haw3}
leads the absorption probability $|\mathcal{A}|^2\sim
\omega^{2l+2}$, which tells us that in the limit
$\omega\rightarrow 0$, the suppression of $|\mathcal{A}|^2$ as the
value of the angular index increases. The dominance of the
absorption probability at the low angular index numbers has also
been shown numerically in \cite{Haw4}.

In Fig.4, we find that for positive $\lambda$, in some range of
frequencies $\omega$, the absorption probability can be negative,
which presents us the super-radiance. This property has not been
observed in the spherical squashed KK black hole in \cite{sq1hw}.

As in \cite{Haw3}, in the low energy limit $BB*\gg i(B*-B)\gg 1$,
we can simplify our (44) to the form
\begin{eqnarray}
|\mathcal{A}_{lm\lambda}|^2=2i(\frac{1}{B}-\frac{1}{B*}) \no
\end{eqnarray}
\begin{eqnarray}
=4\pi [\frac{M\tilde{\omega}(r^2_{\infty}-a^2)}{8\rho_0
r^2_{\infty}}]^{\sqrt{4E_{lm\lambda}+1}}\frac{K*}{A(\rho_+)}\frac{\Gamma^2(2\beta+D_*-2)\Gamma^2(1-\beta)(2-D_*-2\beta)}{\sqrt{E_{lm\lambda}+1/4}\Gamma^2
(\sqrt{E_{lm\lambda}+1/4})\Gamma^2(\beta+D_*-1)\sin^2(\pi(\beta+D_*))}.
\end{eqnarray}
From (27) we learnt that the quantity $2-D_*-2\beta$ is always
positive. Using (19) we have $A(\rho_+)=1-\frac{\rho_0
r^2_{\infty}a^2}{\rho_+(r^4_{\infty}-a^4)-a^4\rho_0}=\frac{(r^2_{\infty}+a^2)(r^2_+-a^2)}{r^2_+(r^2_{\infty}-r^2_-)}$,
which is positive since $r^2_+-a^2=(M-4a^2+\sqrt{M(M-4a^2)})/2>0$.
$\tilde{\omega}>0$ is required to describe the outgoing and
incoming spherical waves in the large distance in (43). The
possibility to make $|\mathcal{A}_{lm\lambda}|^2<0$ is $K_*<0$.
From (21) and (39), $K_*<0$ and $\tilde{\omega}>0$ lead to
\begin{eqnarray}
0\leq \omega\leq \omega_c=\frac{a\lambda
K(\rho_+)^2(r^2_{\infty}+a^2)}{\rho_0
r^3_{\infty}}=\frac{2a\lambda(r^2_{\infty}+a^2)^2}{r_{\infty}(r^2_++a^2)
\sqrt{(r^2_{\infty}+a^2)[(r^2_{\infty}+a^2)^2-Mr^2_{\infty}]}},
\label{wc}
\end{eqnarray}
and
\begin{eqnarray}
\omega\geq\omega_0=\frac{2\lambda(r^2_{\infty}+a^2)^2}{r_{\infty}[(r^2_{\infty}+a^2)^2+Ma^2]}
\frac{\sqrt{(r^2_{\infty}+a^2)[(r^2_{\infty}+a^2)^2-Mr^2_{\infty}]}+Ma}
{\sqrt{(r^2_{\infty}+a^2)[(r^2_{\infty}+a^2)^2-Mr^2_{\infty}]}},\label{w0}
\end{eqnarray}
respectively. The condition for the occurrence of the
super-radiance in this black hole background is
$\omega_0\leq\omega_c$. From (\ref{wc}) and (\ref{w0}), we obtain
the ratio
\begin{eqnarray}
\frac{\omega_0}{\omega_c}=\frac{(r^2_++a^2)\{\sqrt{(r^2_{\infty}+a^2)[(r^2_{\infty}+a^2)^2-Mr^2_{\infty}]}+Ma\}}
{a[(r^2_{\infty}+a^2)^2+Ma^2]}.
\end{eqnarray}
\begin{figure}[ht]
\begin{center}
\includegraphics[width=8.2cm]{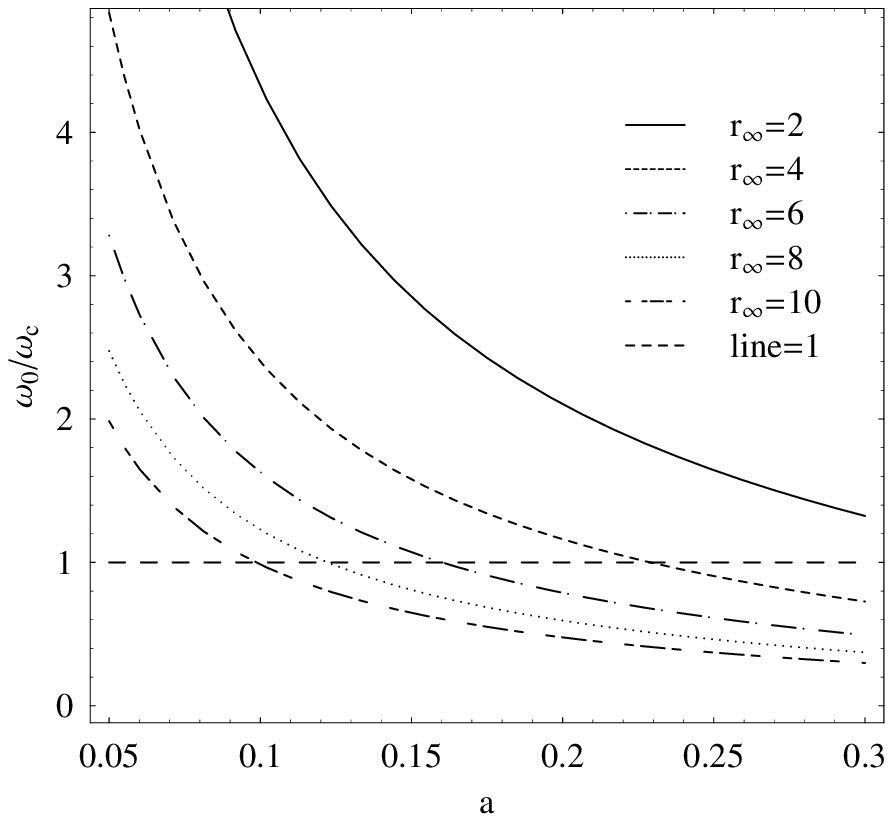}\;\;\;\;\;\;\includegraphics[width=8.2cm]{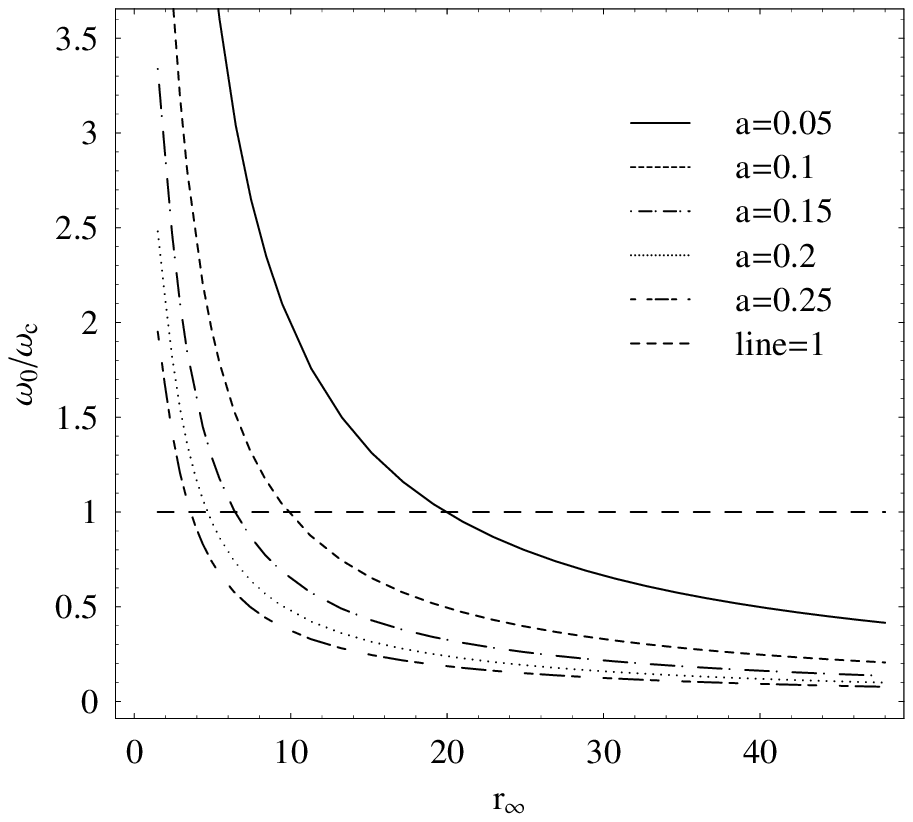}
\caption{Ratio of the value $\omega_0/\omega_c$ with the change of
$a$ and $r_{\infty}$.}
\end{center}
\label{fig5}
\end{figure}
The changes of the ratio with $a$ and $r_{\infty}$ are plotted in
figure (5), which tell us that the larger value of the angular
momentum parameter $a$ leads to the more extended region of the
super-radiance. This has also been observed in \cite{Haw3}. The
increase of $r_{\infty}$ gives the similar result. Moreover, we
observed that for the fixed $r_{\infty}$, there exists a lower bound
of $a$ for the super-radiance to occur. Similarly, for the fixed
$a$, there is also the lowest value of $r_{\infty}$ for the
super-radiance to happen. This is actually shown in Fig.4.

\begin{figure}[ht]
\begin{center}
\includegraphics[width=8.5cm]{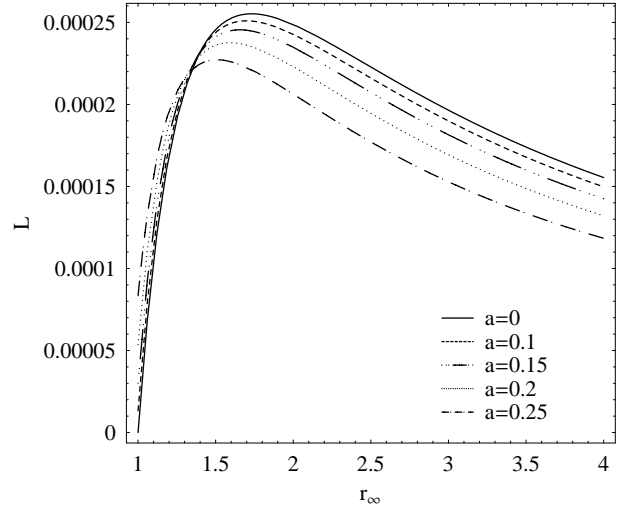}
\caption{The luminosity of Hawking radiation $L$  of scalar
particles propagating in the rotating squashed Kaluza-Klein black
hole spacetime, for $l=0$ and $\lambda=0$ and different
$r_{\infty}$ and $a$.}
\end{center}
\label{fig6}
\end{figure}

Now let us turn to study the luminosity of the Hawking radiation
for the mode $l=0$, $\lambda=0$ which plays a dominant role in the
greybody factor. Performing an analysis similar to that in
\cite{Haw3}, we can rewrite the greybody factor (\ref{GFA}) as
\begin{eqnarray}
|\mathcal{A}_{lm\lambda}|^2&\simeq&
\frac{4\omega^2(\rho_0+\rho_-)}{\rho_0\rho_+}\sqrt{1+\frac{\rho_0}{\rho_+}}
\bigg[\rho_++\frac{a^2}{r^2_{\infty}+a^2}\rho_0\bigg]^3\nonumber\\
&=&\frac{\omega^2
r^5_{\infty}(r^2_++a^2)^3}{r^3_+(r^2_{\infty}+a^2)^2(r^2_{\infty}-r^2_+)}.\label{GFA1}
\end{eqnarray}
For the slowly rotating black hole, when $r_{\infty}$ increases,
the greybody factor decreases. In the limit $a\rightarrow 0$, we
have $\rho_-\rightarrow0$, and then the form of the formula
(\ref{GFA1}) reduces to that in the static squashed Kaluza-Klein
black hole spacetime \cite{sq1hw}. Combining it with equation
(\ref{TH}), the luminosity of the Hawking radiation is given by
\begin{eqnarray}
L&=&\int^{\infty}_0\frac{d\omega}{2\pi}
|\mathcal{A}_{lm\lambda}|^2\frac{\omega}{e^{\;(\omega-\Omega_H\lambda)/T_{H}}-1}\nonumber\\
&\simeq&\frac{1}{1920\pi
\rho^2_+}\bigg(1+\frac{\rho_-}{\rho_0}\bigg)\bigg(1-\frac{\rho_-}{\rho_+}\bigg)^4
\bigg(1+\frac{\rho_0}{\rho_+}\bigg)^{-3/2}\bigg(1+\frac{a^2}{r^2_{\infty}+a^2}\frac{\rho_0}{\rho_+}\bigg)^{-5}
\nonumber\\
&=&\frac{r_+r_{\infty}(r^2_{\infty}-r^2_+)(r^2_{\infty}+a^2)^4(r^2_+-r^2_-)^4}
{480\pi
(r^2_++a^2)^5(r^2_{\infty}-r^2_-)^6}=\frac{\pi^3}{30}GT^4_H,
\label{LHK}
\end{eqnarray}
where $G=\frac{
r^5_{\infty}(r^2_++a^2)^3}{r^3_+(r^2_{\infty}+a^2)^2(r^2_{\infty}-r^2_+)}$.
\begin{figure}[ht]
\begin{center}
\includegraphics[width=8.cm]{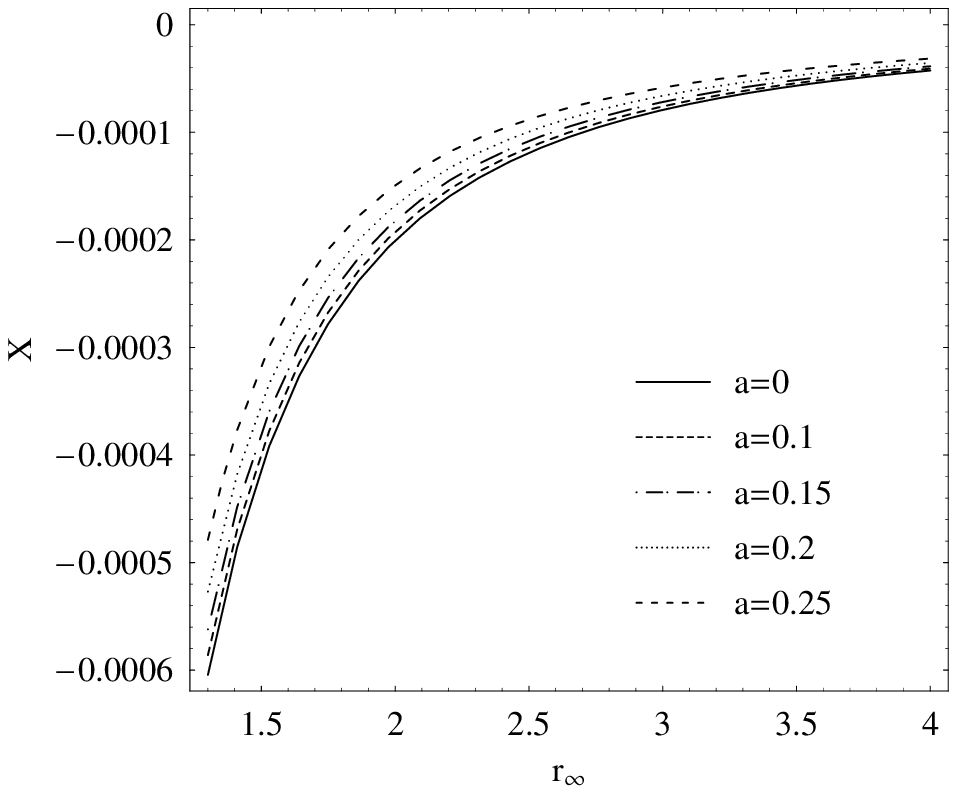}\;\;\;\;\;\;\;\includegraphics[width=8.cm]{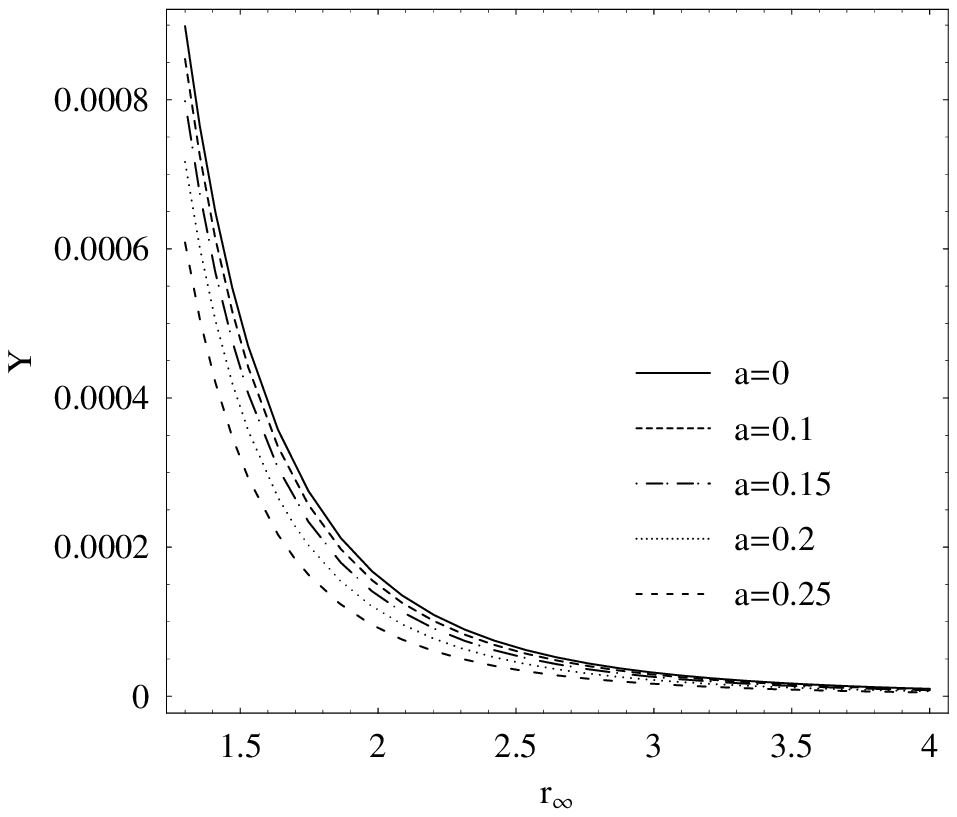}
\\
\includegraphics[width=8cm]{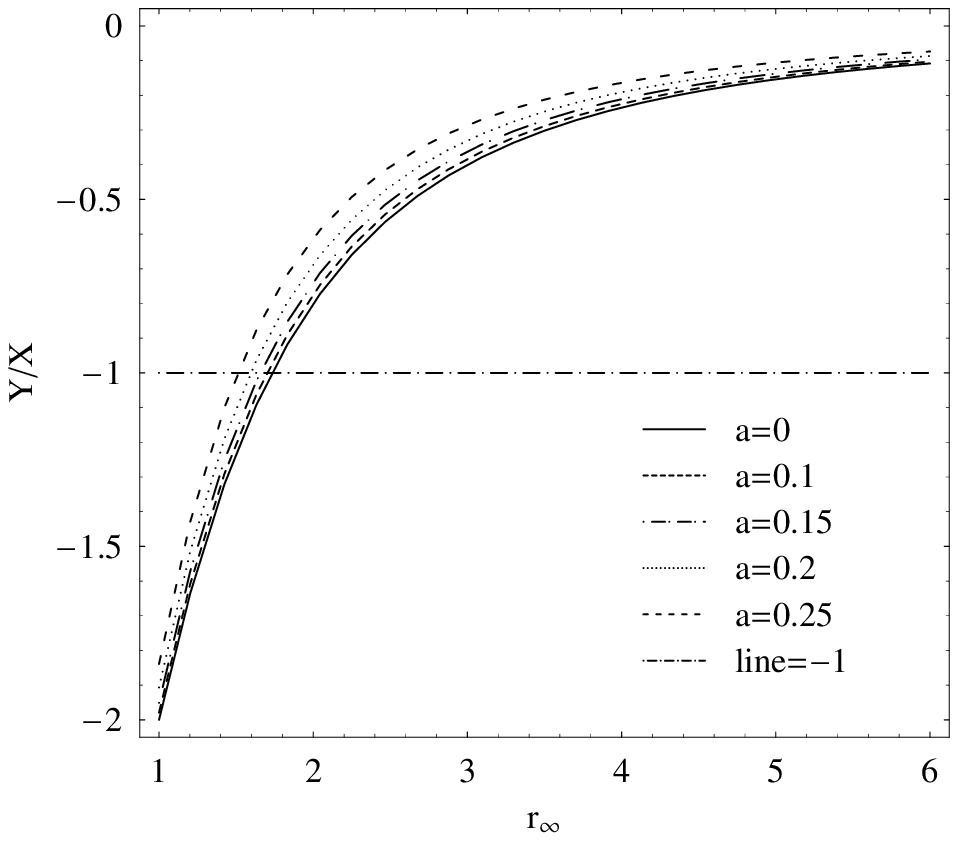}
\caption{The behaviors of $X,Y$ and the ratio $Y/X$ with the
change of $r_{\infty}$ and $a$.}
\end{center}
\label{fig7}
\end{figure}
\begin{table}[h]
\begin{center}
\begin{tabular}[b]{cccccc}
 \hline \hline
 \;\;\;\; $a$ \;\;\;\; & \;\;\;\; 0 \;\;\;\; & \;\;\;\; 0.1\;\;\;\;
 & \;\;\;\; 0.15 \;\;\;\;& \;\;\;\; 0.2 \;\;\;\; & \;\;\;\; 0.25\;\;\;\; \\ \hline
\\
$r_+$& \;\;\;\;\;1\;\;\;\;\;  & \;\;\;\; 0.9899\;\;\;\;\;
 & \;\;\;\;\;0.9770\;\;\;\;& \;\;\;\;\;0.9583\;\;\;\;\; & \;\;\;\; 0.9330\;\;\;\;\;
\\
\\
$r^p_{\infty}$& \;\;\;\;\;1.7321\;\;\;\;\;  & \;\;\;\;
1.6989\;\;\;\;\;
 &\;\;\;\;\;1.6562\;\;\;\;& \;\;\;\;\;1.5938\;\;\;\;\; & \;\;\;\; 1.5092\;\;\;\;\;
 \\
 \\
$\frac{(r^p_{\infty})^2}{r^2_+}$& \;\;\;\;\;3\;\;\;\;\; & \;\;\;\;
2.9454\;\;\;\;\;
 & \;\;\;\;2.8737\;\;\;\;\;& \;\;\;\;\;2.7664\;\;\;\;\; & \;\;\;\; 2.6164\;\;\;\;\;
\\ \hline\hline
\end{tabular}
\caption{The change of $r_+$, $r^p_{\infty}$ and
$(r^p_{\infty})^2/r^2_+$ with different $a$. Here $M=1$. }
\end{center}
\end{table}
In figure 6, we show the dependence of the luminosity of Hawking
radiation on the size of the extra dimension $r_{\infty}$ for
different angular momentum parameters. In the limit
$r_+\rightarrow r_{\infty}$, $L\rightarrow 0$. The limit
$r_+\rightarrow r_{\infty}$ describes that $r_+$ reaches the size
of the fifth dimension at the infinity $r_{\infty}$, which means
that we could effectively obtain a very large four-dimensional
black hole. This limit tells us that the KK black hole has an
infinite horizon radius, as a result the temperature approaches
zero, like the Schwarzschild black hole with infinite radius
\cite{sq1t}, which naturally leads to the vanishing of the
luminosity of the Hawking radiation. When $r_+<r_{\infty}$, from
(50) we know that there is a peak of the luminosity of the Hawking
radiation when $dL/dr_{\infty}=0$. For fixed $M$ but different
$a$, peaks appear at different $r^p_{\infty}$ are listed in table
I. When $r_+<r_{\infty}<r^p_{\infty}$, we observe in Fig.6 that
with the increase of $r_{\infty}$, $L$ increases. This can be
explained by the derivative $dL/dr_{\infty}$ which can be
expressed into $\frac{\pi^3}{30}(X+Y)$, where
$X=T^4_HdG/dr_{\infty}, Y=GdT^4_H/dr_{\infty}$. The variations of
$X,Y$ and $X/Y$ with the change of $r_{\infty}$ are plotted in
Fig.7. $X$ is negative and $Y$ is positive. For small values of
$r_{\infty}$ we have $Y/X <-1$ (i.e $X + Y> 0$), this leads
$dL/dr_{\infty} > 0$. This shows that the extra dimensional effect
is enhanced, which may allow us to detect the extra dimension. For
the larger $r_{\infty}$, we have $Y/X>-1$, (i.e, $X + Y < 0$)
which leads $dL/dr_{\infty} < 0$ and we see in Fig.6 that when
$r_{\infty}>r^p_{\infty}$ the luminosity of Hawking radiation
decreases with the further increase of $r_{\infty}$.

\section{summary}
In this paper, we employed the matching techniques and studied the
low-energy greybody factor and Hawking radiation for a massless
scalar field in the background of a 5-dimensional rotating
squashed KK black hole. We found that the absorption probability
and Hawking radiation contain the imprint of the extra dimensions.
With the inclusion of the rotating parameter $a$, we have observed
richer physics which has not been shown in the spherical squashed
KK black hole background\cite{sq1hw}, such as the super-radiance
and the dependence of the absorption probability and the
luminosity of the Hawking radiation on the rotating parameter. We
found that for an appropriate size of the extra dimension
$r_{\infty}$, the signature of the extra dimension can be enhanced
in the Hawking radiation. This could open a window to detect the
extra dimension.

\begin{acknowledgments}

This work was partially supported by NNSF of China, Ministry of
Education of China and Shanghai Education Commission. R. K. Su's
work was partially supported by the National Basic Research Project
of China. S. B. Chen's work was partially supported by the China
Postdoctoral Science Foundation under Grant No. 20070410685, the
Scientific Research Fund of Hunan Provincial Education Department
Grant No. 07B043 and the National Basic Research Program of China
under Grant No. 2003CB716300.
\end{acknowledgments}

\end{document}